\documentclass[12pt]{article}

\catcode`\@=11

\global\arraycolsep=2pt
\usepackage{amsbsy,amssymb,latexsym,amsfonts,amsmath}
\usepackage{graphicx,color}

\begin{document}
\begin{flushright}
\parbox{4.2cm}
{RUP-16-32}
\end{flushright}

\vspace*{0.7cm}

\begin{center}
{ \Large Interacting scale but non-conformal field theories}
\vspace*{1.5cm}\\
{Yu Nakayama}
\end{center}
\vspace*{1.0cm}
\begin{center}

Department of Physics, Rikkyo University, Toshima, Tokyo 171-8501, Japan

\vspace{3.8cm}
\end{center}

\begin{abstract}
There is a dilemma in constructing interacting scale invariant but not conformal invariant Euclidean field theories. On one hand, scale invariance without conformal invariance seems more generic by requiring only a smaller symmetry. On the other hand, the existence of a non-conserved current with exact scaling dimension $d-1$ in $d$ dimensions seems to require extra fine-tuning. To understand the competition better, we explore some examples  without the reflection positivity.
We show that a theory of elasticity (a.k.a Riva-Cardy theory) coupled with massless fermions in $d=4-\epsilon$ dimensions does not possess an interacting scale invariant fixed point except for unstable (and unphysical) one with an infinite coefficient of compression. We do, however, find interacting scale invariant but non-conformal field theories in gauge fixed versions of the Banks-Zaks fixed points in $d=4$ dimensions. 

\end{abstract}

\thispagestyle{empty} 

\setcounter{page}{0}

\newpage

\section{Introduction}
It is a tantalizing question to find a scale invariant but non-conformal field theory \cite{Nakayama:2013is}. The quest reveals a competition between two genericity arguments. On one hand, conformal symmetry is larger than the mere scale symmetry, so it seems easier to find a scale invariant field theory without conformal invariance. On the other hand, scale invariance without conformal invariance requires the existence of a so-called Virial current \cite{Wess}\cite{Coleman:1970je}, which is non-conserved  but its scaling dimension is exactly $d-1$ in $d$ dimensions.  The latter situation sounds unlikely without extra fine-tuning and indeed this is one of the motivations to assume that the three-dimensional critical Ising model shows conformal invariance \cite{Delamotte:2015aaa}\cite{Paulos:2015jfa}.

Since conformal invariance plays a significant role in solving many critical phenomena, we would like to understand this dilemma better. We want more examples, but many known scale invariant field theories without conformal invariance that  we know are essentially free theories \cite{Jackiw:2011vz}\cite{ElShowk:2011gz}\cite{Dymarsky:2015jia}, and they have huge symmetries to protect the appearance of anomalous dimensions. They have taught us little about how this dilemma is actually avoided. The goal of this paper is to look for non-trivially interacting examples of scale invariant field theories without conformal invariance. 

At this point, we should point out that there has been a culminating evidence (or proof) that scale invariance does imply conformal invariance in two- and four-dimensions if we assume the reflection positivity (together with more subtle assumptions) \cite{Polchinski:1987dy}\cite{Luty:2012ww}\cite{Fortin:2012hn}\cite{Bzowski:2014qja}\cite{Dymarsky:2014zja}\cite{Yonekura:2014tha}\cite{Naseh:2016maw}. In these studies they have explored an elegant interplay between conformal invariance and the structure of the renormalization group.
However, the argument given there appears to have only little to say about the genericity argument given above. In this paper, we abandon the reflection positivity, and focus on the genericity argument more directly. After all, we do know examples of physical systems that do not show the reflection positivity. We encounter some of them in the following discussions. 

The examples we will discuss in this paper are motivated by gauge fixed versions of gauge theories with non-trivial fixed point for the gauge coupling. In the Abelian case, our example may be regarded as a physical model for a theory of elasticity \cite{Landau} (a.k.a Riva-Cardy theory \cite{Riva:2005gd}) coupled with massless fermions. We show that they do not admit any interacting scale invariant fixed point in $d-\epsilon$ dimensions while as a gauge theory, they admit conformal invariant fixed point. In the non-Abelian case, we find interacting scale invariant but non-conformal field theories in gauge fixed versions of the Banks-Zaks fixed points in $d=4$ dimensions. 

The organization of this paper is as follows. In section 2, we review the distinction between scale invariance and conformal invariance. In section 3, we study a theory of elasticity coupled with massless fermions in our attempt to construct scale invariant but not conformal field theory. In section 4, we study gauge fixed versions of non-Abelian gauge theories with massless fundamental fermions. In section 5, we conclude with some discussions.

\section{Scale invariance vs conformal invariance}
In this paper, we study $d$-dimensional Euclidean invariant field theories. The Euclidean invariance requires the existence of a symmetric conserved energy-momentum tensor $T_{\mu\nu} = T_{\nu\mu}$ such that
\begin{align}
\partial^\mu T_{\mu\nu} =  0  \ .
\end{align}

The scale invariance requires that the trace of the energy-momentum tensor is given by a divergence of a certain (non-conserved) current $J_\mu$:
\begin{align}
T^\mu_{\mu} = \partial^\mu J_\mu \ ,
\end{align}
which is often called the Virial current.
The scaling dimension of the energy-momentum tensor is exactly $d$.\footnote{Under the dilatation, it may mix with lower dimensional operators with a triangular mixing matrix, but the diagonal part must have dimension $d$.} Accordingly, the scaling dimension of the Virial current must be exactly $d-1$.

The conformal invariance demands that we may improve the energy-momentum tensor so that it becomes traceless:
\begin{align}
\tilde{T}^\mu_{\mu} = 0 \ .
\end{align}
For this to be possible, the trace of the Virial current must be given by 
\begin{align}
J_\mu = \partial_\mu L + \partial^\nu L_{\mu\nu} \ 
\end{align}
for certain local operators $L$ and $L_{\mu\nu}$. 
If this is not the case, the theory is invariant only under scale symmetry rather than full conformal symmetry.

One important consequence of the conformal symmetry is a restriction of two-point functions for vector primary operators $O_\mu$. By primary operators, we mean that they are not written as derivatives of some other local operators.\footnote{In conformal field theories, we may define them being annihilated by special conformal transformation, but such a definition is not available without conformal invariance. The notion here may be a little imprecise without defining what we mean by ``local operators". We will not concern ourselves with these problems by working in perturbative field theories with the Lagrangian description in which the fundamental degrees of freedom is explicit.} 
In the momentum space the conformal invariance requires 
\begin{align}
\langle O_\mu(p) O_\nu(q) \rangle = (2\pi)^d\delta^{(d)}(p+q) p^{2(\Delta-d+1)} \left(p^2 \delta_{\mu\nu} - \frac{2\Delta-d}{\Delta-1}p_\mu p_\nu \right) \label{tpf}
\end{align}
in $d$ dimensions, where $\Delta$ is the conformal dimension of the operator $O_\mu(x)$.
Note that when $O_\mu$ is conserved, the conformal dimension must satisfy $\Delta = d-1$ in $d$-dimensions.\footnote{In $d=4$ with $\Delta=3$, \eqref{tpf} acquire a logarithmic correction due to the trace anomaly.}

\section{Riva-Cardy model with massless fermions}
Let us begin with one of the simplest examples of scale invariant but not conformal invariant field theories.
A theory of elasticity is described by a model with the Euclidean action
\begin{align}
 S = \int d^dx \frac{1}{4}(\partial_\mu v_\nu - \partial_\nu v_\mu)^2 + \frac{\alpha^{-1}}{2} (\partial^\mu v_\mu)^2 \ . \label{rc}
\end{align}
Here $v_\mu$ is what is called the displacement vector \cite{Landau}.

The momentum space two-point function of $v_\mu$ is given by
\begin{align}
\langle v_\mu(p) v_\nu(q) \rangle = (2\pi)^d \delta(p+q) \left(\frac{\delta_{\mu\nu}}{p^2} - (1-\alpha) \frac{p_\mu p_\nu}{p^2} \right) \ . \label{two}
\end{align}
As discussed in \cite{ElShowk:2011gz}, the theory is scale invariant but not conformal invariant except for the special value of $\alpha = \frac{d}{d-4}$. One way to see this is to note that \eqref{two} does not satisfy the conformal Ward identity by assuming $v_\mu$ is a vector primary field (compare \eqref{two} with \eqref{tpf}).\footnote{This condition may be relaxed by declaring $v_\mu$ is a descendant in $d=2$ dimensions \cite{Nakayama:2016dby}.} Another way is to compute the trace of the energy-momentum tensor
\begin{align}
T^\mu_{\mu} = \left(\frac{d}{2}-2\right) (\partial_\mu v_\nu \partial^\mu v^\nu - \partial_\mu v_\nu \partial^\nu v^\mu) + \alpha^{-1}\left((2-d)v_\mu\partial^\mu\partial^\nu v_\nu - \frac{d}{2}(\partial_\mu v^\mu)^2 \right) \ ,
\end{align}
and check that it is not possible to improve it to be traceless, which requires that 
\begin{align}
T^\mu_{\mu} = a \partial^\nu \partial_\nu (v^\mu v_\mu) + b \partial^\mu \partial^\nu (v_\mu v_\nu) \ 
\end{align}
for certain $a$ and $b$ up to the use of the equation of motion. Again this is only possible when $\alpha = \frac{d}{d-4}$.

One may also regard the theory as a twisted deformation of a $O(4)$ symmetric scalar field theory as discussed in \cite{Nakayama:2016ydc}. Suppose we begin with massless free scalars $v^i$ with $O(4)$ global symmetry, which is conformal invariant. We now identify this $O(4)$ with the Euclidean rotation by using $\delta^{\mu}_i$ as a twist. This gives rise to the action \eqref{rc} at $\alpha = 1$ with $v^i$ now being identified as a  Euclidean vector $v_\mu$. Then we can deform the theory by changing $\alpha$, which is what we name the twisted scalar deformations in \cite{Nakayama:2016ydc} proposed as a way to obtain theories with scale invariance but without conformal invariance. After the deformation, one cannot undone the twist.

We add some massless matters to look for a non-trivially interacting  scale invariant fixed point. For instance, let us add $N_f$ massless (Dirac) fermions with the ``minimal" coupling
\begin{align}
S = \int d^dx \bar{\psi}_i (\partial_\mu - ie v_\mu) \gamma^\mu \psi_i \ . \label{minimal}
\end{align}
At this point, we could have added the self-couplings of the displacement vector:
\begin{align}
S = \int d^dx m^2 (v_\mu v^\mu) + \lambda (v_\mu v^\mu)^2 + \zeta (\partial_\mu v^\mu) (v_\nu v^\nu) \ , \label{nrc}
\end{align}
but it turns out that $m^2 = \lambda = \zeta = 0$ can be preserved in the renormalization group flow, so we set $m^2 = \lambda = \zeta = 0$.

As a theory of elasticity, it is reasonable to assume that the free energy has a symmetry under the constant shift of the displacement vector $v_\mu \to v_\mu + c_\mu$, which excludes all the terms in \eqref{nrc}. The minimal coupling \eqref{minimal} is still allowed by assigning $\psi \to \psi e^{ie x^\mu c_\mu} $. This symmetry is spontaneously broken and the displacement vector may be regarded as a Nambu-Goldstone boson with this respect.

An alternative viewpoint of this model is to regard it as a gauged fixed version of QED in $d=4-\epsilon$ dimensions. It is known that {\it as a gauge theory}, the theory has an interacting {\it conformal} fixed point for sufficiently large number of $N_f$ when $\epsilon$ is small. The question we would like to address here is whether it shows a scale invariant but not necessarily conformal invariant fixed point as a gauge fixed theory defined by the action \eqref{rc} and \eqref{minimal}.

To look for an interacting scale invariant fixed point, we first look at the beta function for $e$ and $\alpha$, which can be computed in perturbation theory at one-loop order as
\begin{align}
\beta_e &= \mu \frac{\partial e}{\partial \mu} = -\epsilon e + \frac{N_f}{12\pi^2} e^3  + O(e^4) \cr 
\beta_{\alpha} &= \mu \frac{\partial \alpha}{\partial \mu}  = -\gamma_3\alpha  = -2\alpha \frac{N_f}{6\pi^2}e^2 + O(e^4) \ ,
\end{align}
where $\gamma_3$ is the anomalous dimensions of $v_\mu$.
Note that $\beta_e$ does not depend on $\alpha$, which is true all orders in perturbation theory. Note also that we define the renormalized coupling constant $\alpha$ as a {\it dimensionless} ratio of the kinetic term in any $d$ dimensions, so there is no classical beta function unlike in $\beta_{e}$. As long as $\epsilon/N_f$ is sufficiently small, we see that $e$ has a non-trivial fixed point, but we also see that within perturbation theory there is no non-trivial renormalization group fixed point for $\alpha$ except at $\alpha=0$, which is singular as a theory of elasticity as it describes a material with an infinite coefficient of compression \cite{Landau}.

Therefore we claim that the theory of elasticity coupled with massless Dirac fermions do not show any scale invariant renormalization group fixed point.  Of course, this does not contradict with the fact that as a gauge theory, it has a non-trivial conformal fixed point because the term containing gauge fixing parameter $\alpha$ is BRST exact and it has no physical effects as a gauge theory.

To understand the physical picture better, let us compute the two-point function of the displacement vector $v_\mu$. For this purpose, it is convenient to regard it as a gauge fixed version of QED.
As long as $e$ has a non-trivial fixed point, the two-point function must be exactly given by
\begin{align}
\langle v_\mu(p) v_\nu(q) \rangle = (2\pi)^d \delta(p+q) \left(\frac{\delta_{\mu\nu} - \frac{p_\mu p_\nu}{p^2}}{p^{d-2}}  + \alpha \frac{p_\mu p_\nu}{p^2} \right) 
\end{align}
because the Bianchi identity tells that $v_\mu$ must possess an exact scaling dimension $1$ in any $d$-dimensions \cite{DiPietro:2015taa} and there is no renormalization for the transverse part of $v_\mu$.  We conclude that there is no non-trivial scale invariant, let alone conformal, fixed point for the theory of elasticity with minimally coupled massless Dirac fermions. 

As a gauge theory, it may be ammusing to note that if we fix the gauge in the renormalizable $\xi$ gauge, the theory in $d=4-\epsilon$ dimension never shows scale invariance except in the Landau gauge $\alpha=0$ even though the BRST invariant correlation functions show conformal invariance.\footnote{The special feature of the Landau gague was also emphasized in \cite{Chester:2016ref}} The absolute value of the gauge fixing parameter becomes effectively larger in the infrared, so the Landau gauge is an unstable fixed point.

Let us discuss some properties of the Virial current at the scale invariant fixed point of $\alpha = 0$. Since the original action is singular at $\alpha = 0$, we alternatively use 
\begin{align}
 S = \int d^dx \frac{1}{4}(\partial_\mu v_\nu - \partial_\nu v_\mu)^2 + iB\partial^\mu A_\mu + i \partial^\mu \bar{c} \partial_\mu c \ , \label{rc}
\end{align}
where $B$ is the Legendre multiplier field and $c$ and $\bar{c}$ are decoupled ghost and anti-ghost, which we will use to make the BRST invariance manifest. 
The Virial current is given by
\begin{align}
J_\mu \propto [Q_{BRST}, i\bar{c} A_\mu] = -i\bar{c} \partial_\mu c + iBA_\mu \ .
\end{align}
Here $[Q_{BRST}, A_\mu] = \partial_\mu c$ and $\{Q_{BRST}, \bar{c} \} = B$ as usual.
To compute the scaling dimension of the Virial current, we realize
\begin{align}
\Delta(J_\mu) = \Delta(Q_{BRST}) + \Delta(\bar{c}) + \Delta(A_\mu) 
\end{align}
because anti-ghost completely decouples in this theory. Now, we recall 
$\Delta (A_\mu) = 1$ at the fixed point, so from $[Q_{BRST}, A_\mu] = \partial_\mu c$, we require $\Delta(Q_{BRST}) = \Delta(c) = \Delta(\bar{c}) = \frac{d-2}{2}$. Therefore, we obtain $\Delta(J_\mu) = d-1$ exactly as claimed in section 2.

Finally, we may generalize our discussions with other massless matters such as scalars. Once we introduce the shift symmetry $v_\mu \to v_\mu + c_\mu$, the effective action is essentially described by the gauge fixed version of QED in $d=4-\epsilon$ dimensions with minimally coupled matters. Therefore the above argument applies with no change in details. In particular, we do not find any interacting scale invariant, let alone conformally invariant, fixed point.

\section{Gauge fixed version of Banks-Zaks theory}
Since a theory of elasiticity coupled with massless matters does not admit any interacting scale invariant but not necessarily conformal invariant fixed point in $d=4-\epsilon$ dimensions, we look for more elaborate examples. For this purpose, we consider gauge fixed versions of non-Abelian gauge theories in $d=4$ dimensions.

Concretely, we study the $SU(N)$ gauge theory with $N_f$ massless fermions in the fundamental representation and fix the gauge in the renormalizable $\xi$ gauge. Including the ghost, the entire action is given by
\begin{align}
 S = \int d^4x\frac{1}{g^2} \left(\frac{1}{4}(F_{\mu\nu}^a)^2 + \frac{1}{2\alpha} (\partial^\mu A_\mu^a)^2 \right) + \bar{\psi}_i D_\mu \gamma^\mu \psi_i  + i\bar{c}^a\partial^\mu D_\mu c^a \ .
\end{align}

The beta functions for the gauge coupling $g$ and the gauge fixing parameter $\alpha$ are computed as \cite{Caswell:1974gg}
\begin{align}
\frac{\beta_g}{g}  &= -\frac{g^2}{16\pi^2} \left(\frac{11}{3}C_2 - \frac{2}{3} n_f \right) - \left(\frac{g^2}{16\pi^2}\right)^2\left(\frac{34}{3}C_2^2 - 2C_F n_f - \frac{10}{3} C_2 n_f \right) \cr
\beta_{\alpha} &= -\alpha \gamma_3  = -\alpha \frac{g^2}{16\pi^2} \left( \left(-\frac{5}{3} -\frac{(1-\alpha)}{2} \right)C_2 + \frac{2}{3}n_f \right) \cr
&-\alpha \left(\frac{g^2}{16\pi^2}\right)^2 \left(\left(\frac{(1-\alpha)^2}{4} -\frac{15}{8}(1-\alpha) -\frac{23}{4}\right)C_2^2 + 2C_F n_f + \frac{5}{2} C_2 n_f \right) \ 
\end{align}
at the two-loop order, where $C_F = \frac{N_c^2-1}{2N_c}$ and $C_2= N_c$. 

Let us discuss the structure of the renormalization group flow near the Banks-Zaks fixed point \cite{Banks:1981nn}, i.e. $N_f^*<N_f \le \frac{11}{2} C_2$ (while the critical number of flavor $N_f^*$ is currently unknown in a precise manner) with $g=g^*$ where $\beta_g(g^*) = 0$. When the fixed point gauge coupling is sufficiently small, we see that $\alpha$ has a fixed point at $\alpha = 0$ and $\alpha = \alpha^* \sim  -\frac{4 N_f}{3C_2} + \frac{13}{3}  < 0$. The former fixed point is unstable while the latter is stable toward the infrared. See Figure 1 for the renormalization group flow in the case of $N_c =3$ and $N_f=16$.

\begin{figure}[htbp]
	\begin{center}
  \includegraphics[width=9.0cm,clip]{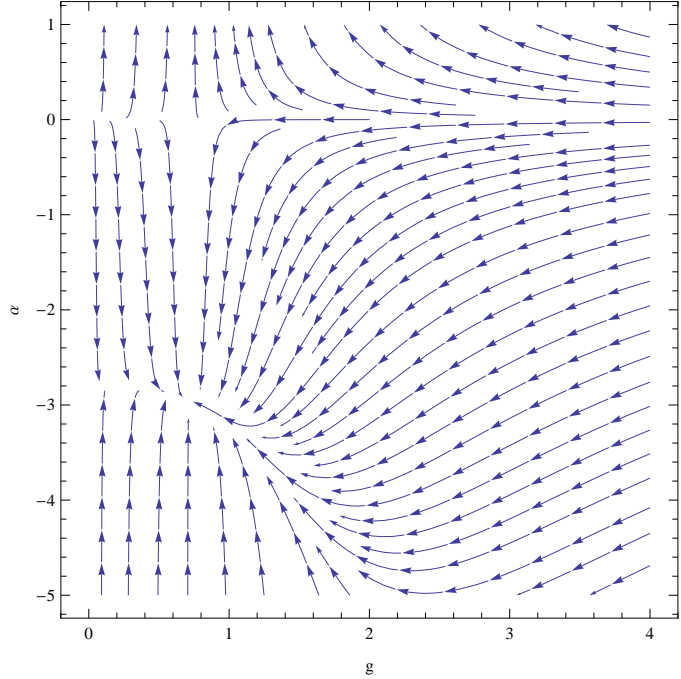}
  \end{center}
  \caption{Renormalization group flow of the gauge coupling $g$ and the gauge fixing parameter $\alpha$ in $SU(3)$ gauge theories with $N_f=16$ fermions.}
  \label{fig:charge_3}
\end{figure}

There is a difference between the two fixed points for $\alpha$. The fixed point with $\alpha = 0$ corresponds to the Landau gauge, and the two-point function of the vector potential behaves as 
\begin{align}
\langle A_\mu^a(p) A_\nu^b (q) \rangle = (2\pi)^d \delta(p+q) \frac{\delta^{ab}}{p^{2-2\gamma_3}} \left(\delta_{\mu\nu} - \frac{p_\mu p_\nu}{p^2} \right)
\end{align}
with non-zero anomalous dimension $\gamma_3$. On the other hand, at the fixed point with  non-zero $\alpha$, the two-point function of the vector potential behaves as 
\begin{align}
\langle A_\mu^a(p) A_\nu^b (q) \rangle = (2\pi)^d \delta(p+q) \frac{\delta^{ab}}{p^{2}} \left(\delta_{\mu\nu} - (1-\alpha^*) \frac{p_\mu p_\nu}{p^2}  \right)
\end{align}
with the zero anomalous dimension. In both cases, the two-point function is not compatible with the special conformal transformation if we regard the vector potential $A_\mu^a$ as a primary operator. This is not in contradiction with the fact that the Banks-Zaks fixed point as a gauge theory is a conformal field theory, at least in the perturbative regime, because the vector potential is not BRST invariant and the gauge invariant operators cannot be constructed just by taking derivatives of them unlike in QED.

To understand how we get the scale invariance without conformal invariance better, it is instructive to study the properties of the trace of the energy-momentum tensor.
In $d=4-\epsilon$ dimensions, the trace of the bare energy-momentum tensor is computed as
\begin{align}
T^\mu_{\mu} =\frac{\epsilon}{g^2} \left(\frac{1}{4}(F_{\mu\nu}^a)^2+ \frac{1}{2\alpha} (\partial^\mu A_\mu^a)^2 \right)  + \partial^\mu J_\mu \ ,
\end{align}
where the bare Virial current is given by
\begin{align}
J_\mu = (2-\epsilon) ( \alpha^{-1} A_\mu^a \partial^\nu A_\nu^a + i\bar{c}D_\mu c) \ .
\end{align}

In order to obtain the trace of the renormalized energy-momentum tensor in $d=4$ dimensions, we take the $1/\epsilon$ poles in the coupling constant $g$ and $\alpha$ so that
\begin{align}
T^\mu_{\mu} = -\frac{\beta_g}{2g^3}( F^a_{\mu\nu})^2 - \left(\frac{\beta_g}{g^3 \alpha} + \frac{\beta_\alpha}{2g^2 \alpha^2} \right)(\partial^\mu A^a_\mu)^2 + \partial^\mu J_\mu \ .
\end{align}
There are also $1/\epsilon$ poles from the wavefunction renormalization but it is proportional to $\frac{\delta S}{\delta \phi}$ and they vanish if we use the equations of motion. In principle, the Virial current $J_\mu$ may be renormalized but it cannot affect the beta functions for $g$ and $\alpha$.\footnote{In the case of the Landau gauge, if we introduce the auxiliary field $B$ in the Lagrangian so that $L =  \frac{1}{4g^2} (F_{\mu\nu}^a)^2+ B^a \partial^\mu A^a_\mu$, then the trace of the energy-momentum tensor becomes $T^\mu_\mu = \partial^\mu B^a A_\mu^a$ at $\beta(g)=0$. We do not have to tune the anomalous dimension $\gamma_3$ to obtain scale invariance.}

This analysis again tells us that in order to retain scale invariance in a gauge fixed version of the multi-flavor QCD, we need to demand $\beta_g = \beta_\alpha = 0$. Furthermore, the renormalization of the Virial current operator $J_\mu$ is  $O(g^2)$, if any, so in perturbation theory, the fixed point is only scale invariant but not conformal invariant. However, we also notice that the Virial current operator is BRST exact 
\begin{align}
J_\mu \propto \left\{Q_{B}, i\bar{c}A_\mu \right\} \ 
\end{align}
as it should, so within the BRST cohomology, one may realize the conformal symmetry.

In fact, as shown in \cite{Collins:1976yq}, by using the BRST symmetry, one may argue that after the wavefunction renormalization, the divergence of the Virial current operator $\partial^\mu J_\mu$ is finite to all orders in perturbation theory with dimensional regularization, and hence it does not require any further renormalization.\footnote{For readers' convenience, we give a sketch of their argument in Appendix.} 
This means that the Virial current does not acquire any anomalous dimensions at the scale invariant but not conformal fixed point. 

\section{Discussions}
In this paper, we have shown some examples of interacting field theories with scale invariance without conformal invariance. Although examples may not represent the generic features, these are rare and unexplored ones, so let us revisit our original questions, in particular, the (non-)genericity of scale invariance without conformal invariance in view of the  fine-tuning.

If we regard our examples as gauge theories, in order to obtain the conformal invariant fixed point, we have to find the zero of the beta function for the gauge coupling constant. Since there is no candidate for the gauge invariant Virial current in perturbation theory, demanding scale invariance is equivalent to demanding conformal invariance. We need to solve one equation with one parameter.

On the other hand, as a gauge fixed theory, we may find a scale invariant but not conformal fixed point because there is a candidate for the Virial current operator. At the same time, we have additional gauge fixing parameter as a coupling constant. In order to reach the scale invariant fixed point, we have to solve two equations with two variables. If we further demand conformal invariance, we have to add one more constraint, which generically admits no solutions. Indeed in perturbation theory, we did not find any conformal invariant fixed point.

From what we have learned in these examples, we actually see that scale invariance is more generic than conformal invariance once we accept the possibility of the Virial current operator. The last question, then, is if the existence of the Virial current operator itself is natural or not. In our examples, once the theory has vanishing beta functions for the gauge coupling constant and the gauge fixing parameter, the Virial current does have an exact scaling dimension of $d-1$. In our examples, this ``miracle" may be attributed to the underlying BRST symmetry. It is an interesting question to see if the ``miracle" can occur without a hidden symmetry. 

With this respect, it may be important to revisit the holographic examples studied in \cite{Nakayama:2009qu}\cite{Nakayama:2009fe}\cite{Nakayama:2010zz}\cite{Nakayama:2016ydc}\cite{Nakayama:2016xzs}. In these models, we have the non-conserved vector operator whose scaling dimension is exactly $d-1$. From the effective field theory viewpoint, this is certainly fine-tuning by a judicious choice of the effective action, but the recent model proposed in \cite{Nakayama:2016ydc}\cite{Nakayama:2016xzs} has an explicit M-theory realization, so one may expect a field theory reason why such fine-tuning is naturally realized.

Finally, our analysis may be important in our studies of conformal field theories realized by gauge theories. We have shown that the covariant gauge fixing we often use is not consistent with the conformal symmetry, so if we approach them by using the gauge fixed path integral, then the beauty of the conformal symmetry is not manifest in its computation.\footnote{Such subtleties should be discussed with care when we use localization or Hamiltonian truncation to study the gauge theory, in which conformal symmetry plays a significant role.} 
One may still preserve scale invariance by a judicious choice of the gauge parameter. The use of such a gauge should be seen in the future. 

\section*{Acknowledgement}
When we got a bus ride from the Beijing Friendship hotel to the Tsinghua University, S.~Rychkov asked me if the Riva-Cardy theory coupled with massless fermions show scale invariance without conformal invariance, and I said I believe so, but he said he does not think so. It turns out he is right (except at one point!). I thank him for the stimulating discussions. This work is supported in part by Rikkyo University Special Fund for Research.

\appendix
\section{Non-renormalization of the Virial current operator}
We review the argument in \cite{Collins:1976yq} that the Virial current operator is not renormalized to all orders in perturbation theory due to the BRST invariance. 

Let us first recall what we mean by the renormalizablity of a gauge theory with the BRST symmetry. To present the BRST invariance in a compact form, let us introduce the source terms $K^I$ for the BRST transforms of the fundamental fields $\phi_I = \{A_\mu, \bar{c},c, \psi ...\}$.
\begin{align}
S[\phi_I,K^I] = S_0 + \int d^4x K^I [Q_{BRST}, \phi_I] 
\end{align}
Since BRST transformation is nilpotent,\footnote{This is true for the source terms off shell if we use the auxiliary $B$ field, but even without it, the following argument does not change much. In particular, in \cite{Collins:1976yq}, they did not introduce the source term for $\partial^\mu A_\mu$ for their argument reproduced below.} $S[\phi_I,K^I]$ is BRST invariant, and so must be the (regularized but yet-to-be-renormalized) 1PI effective action $\Gamma_0[\phi_I,K^I]$. In other words, the bare Zinn-Justin equation
\begin{align}
\int d^4x \frac{\delta \Gamma_0}{\delta \phi_I} \frac{\delta \Gamma_0}{\delta K^I} = 0 
\end{align}
must hold (by noting that derivative with respect to $K^I$ is nothing but the BRST transform of $\phi_I$). 

This equation tells us what kind of divergence occurs in the gauge theory. Let us decompose the local part of the 1PI effective action $\Gamma_0 = S_0 + \Gamma_{\mathrm{div}} + \text{non local}$, so that the local divergent part of the 1PI effective action must satisfy
\begin{align}
\int d^4x  \frac{\delta S_0}{\delta \phi_I} \frac{\delta \Gamma_{\mathrm{div}}}{\delta K^I} 
+  \frac{\delta \Gamma_{\mathrm{div}}}{\delta \phi_I} \frac{\delta S_0}{\delta K^I}= 0 \ ,
\end{align}
which is known as the renormalization equation and dictates what kind of divergence occurs.

Without giving a proof here, the solution of this algebraic renormalization equation, if restricted to the local functional whose canonical dimension is equal or less than four, takes the same form as the variation of $S_0$ under the wavefunction renormalization and the renormalization of the coupling constant.
This is the claim of the renormalizablity of the gauge theory just by replacing  $(A^\mu \to Z_3^{1/2} A^\mu_R$. $c \to \tilde{Z}^{1/2}c_R$, $\psi = Z_2^{1/2} \psi_R$, $g \to g_R Z_1^{-1} Z_2^{-1/2} Z_3^{-1/2}$, $\xi \to \xi_R Z_3^{-1}$).  
We note that the source $K_I$ are multiplicatively renormalized so that the Zinn-Justin equation  takes the same form
\begin{align}
\int d^4x \frac{\delta \Gamma_R}{\delta \phi_{IR}} \frac{\delta \Gamma_R}{\delta K^I_R} = (Z_3\tilde{Z})^{1/2} \int d^4x \frac{\delta \Gamma_0}{\delta \phi_I} \frac{\delta \Gamma_0}{\delta K^I} = 0 \label{renorm}
\end{align}
in terms of the renormalized quantities.

Let us generalize the above argument by adding the other operator and its source in the action in order to discuss the constraint on the renormalizaiton of $O$:
\begin{align}
S[\phi_I,K^I,J] = S_0 + \int d^4x K^I [Q_{BRST}, \phi_I] + JO \ .
\end{align}
As a warm-up exercise, let us first consider the case in which $O$ is a gauge invariant operator (such as $(\mathrm{Tr} F^{\mu\nu} F_{\mu\nu})^2$). 
In this case, all the derivations above do not change and we have the equation
\begin{align}
\int d^4x \frac{\delta \Gamma_0[J]}{\delta \phi_I} \frac{\delta \Gamma_0[J]}{\delta K^I} = 0 \ .
\end{align}
In particular, if the added operator has dimension less than four, this equation shows that all the divergence of the correlation functions involving $O$ can be removed by the wavefunction renormalization and the charge renormalization. Essentially, it means that the gauge invariant operator only mixes with gauge invariant operators.\footnote{If, on the other hand, the added operator has dimension greater than four, we have to redo the analysis of the renormalization equation.}

For our actual purpose, let us take $J O = \tau \partial^\mu J_\mu$ with the Virial current operator $J_\mu$ and the dilaton $\tau$. This operator is almost (or on-shell) BRST invariant: $[Q_{BRST}, \partial^\mu J_\mu] \propto \partial^\mu (A_\mu \partial^\nu D_\nu c)$. In the 1PI effective action, one may replace the left hand side of the ghost equations of motion $\partial^\mu D_\mu c= 0$ by derivative with respect to $\bar{c}$, so the bare Zinn-Junstin equation now looks like
\begin{align}
\int d^4x \frac{\delta \Gamma_0[\tau]}{\delta \phi_I} \frac{\delta \Gamma_0[\tau]}{\delta K^I}  =  \int d^4x \frac{\tau}{\alpha_0} \partial^\mu (A^\mu \frac{\delta \Gamma_0[\tau]}{\delta \bar{c}}) . 
\end{align}
Or, by taking the derivative with respect to $\tau$ and setting $\tau = 0$, 
\begin{align}
\frac{\delta}{\delta \tau} \int d^4x \frac{\delta \Gamma_0[\tau]}{\delta \phi_I} \frac{\delta \Gamma_0[\tau]}{\delta K^I} |_{\tau =0} = \frac{1}{\alpha_0} \partial^\mu (A^\mu \frac{\delta \Gamma_0}{\delta \bar{c}}) . 
\end{align}

A crucial observation is that if we do the wavefunction renormalization and coupling constant renormalization that we did in \eqref{renorm} and divide by the common factor $(Z_3\tilde{Z})^{1/2}$, then the right hand side is finite because everything is written in the renormalized quantities. The left hand side may be still divergent, but it must satisfy the same renormalization equation
\begin{align}
\int d^4 x \frac{\delta S_0}{\delta \phi_I} \frac{\delta \tilde{\Gamma}_{\mathrm{div}}}{\delta K^I} 
+  \frac{\delta \tilde{\Gamma}_{\mathrm{div}}}{\delta \phi_I} \frac{\delta S_0}{\delta K^I}= 0 \ ,
\end{align}
where $\tilde{\Gamma} = \frac{\delta \Gamma[\tau]}{\delta \tau}|_{\tau=0}$ is the generating functions of the 1PI diagrams with one Virial current operator insertion. 
The possible divergence here means that we may need additional renormalization for the Virial current operator even after the wavefunction renormalizaiton and the coupling constant renormalizaiton. However, the form of the renormalization equation implies that the divergence must be removed by additional wavefunction renormalization and the coupling constant renormalization. In other words, the divergence of the Virial current operator (i.e. the trace of the energy-momentum tensor) may be renormalized by the gauge invariant operators such as $\mathrm{Tr}F^{\mu\nu} F_{\mu\nu}$ or  the redundant operators (that are proportional to the equations of motion). On the other hand, all these operators are not the form of $\partial^\mu J_\mu$, so actually, the insertion of the Virial current operator does not require any renormalization at all at the scale invariant fixed point. This is the claim made in \cite{Collins:1976yq}.

\end{document}